\documentclass{article}
\usepackage[utf8]{inputenc}
\usepackage{graphicx}
\usepackage{color}
\usepackage[toc,page]{appendix}
\usepackage{amsmath}
\usepackage{float}
\usepackage{authblk}
\usepackage[sort, numbers]{natbib}
\usepackage[sort, numbers]{natbib}
\usepackage{booktabs}
\usepackage{xcolor}

\title{
Contagion mean field model for transport
in urban traffic networks
}
\author[1,2]{Arturo Berrones Santos}
\author[3]{Gerardo Palafox Castillo}
\author[4]{Sareé González Huesca}
\author[4]{Carlos Alberto Aldana Sandoval}
\affil[1]{Universidad Autónoma de Nuevo León, 
Posgrado en Ingeniería de Sistemas, 
Facultad de Ingeniería Mecánica y Eléctrica
(Graduate Program in Systems Engineering, 
Autonomous University of Nuevo León), Mexico}
\affil[2]{
Universidad Autónoma de Nuevo León,
Posgrado en Ciencias con Orientación en Matemáticas,
Facultad de Ciencias Físico-Matemáticas
(Graduate Program in Mathematical Sciences,
Autonomous University of Nuevo León), Mexico
}
\affil[3]{
Universidad Autónoma de Nuevo León,
Facultad de Ciencias Físico-Matemáticas
(Autonomous University of Nuevo León), Mexico
}
\affil[4]{
Universidad Autónoma de Nuevo León,
Programa de Verano de la Investigación Científica y Tecnológica, 
Facultad de Ingeniería Mecánica y Eléctrica
(Summer Program for Scientific and Technological Research,
Autonomous University of Nuevo León), Mexico
}

\date{}

\begin{document}

\maketitle

\begin{abstract}
  Theoretical arguments and empirical evidence 
  for the emergence of macroscopic epidemic type
  behavior, in the form of 
  Susceptible-Infected-Susceptible
  (SIS) or Susceptible-Infected-Recovered (SIR)
  processes
  in urban traffic congestion from
  microscopic network flows is given.
  Moreover, it's shown that the emergence of 
  SIS/SIR implies a 
  relationship between traffic flow and density, 
  which 
  is consistent with observations of 
  the so called \emph{Fundamental Diagram of Traffic} (FDT), 
  which
  is a characteristic signature of vehicle movement phenomena that spans 
  multiple scales.
  Our results 
  put in more firm grounds recent findings 
  that indicate that traffic congestion at the aggregate level 
  can be modeled by simple contagion dynamics. 
\end{abstract}

\section{Introduction} \label{intro}

\begin{figure}[ht]
\begin{center}
   \includegraphics[width=13.0cm]{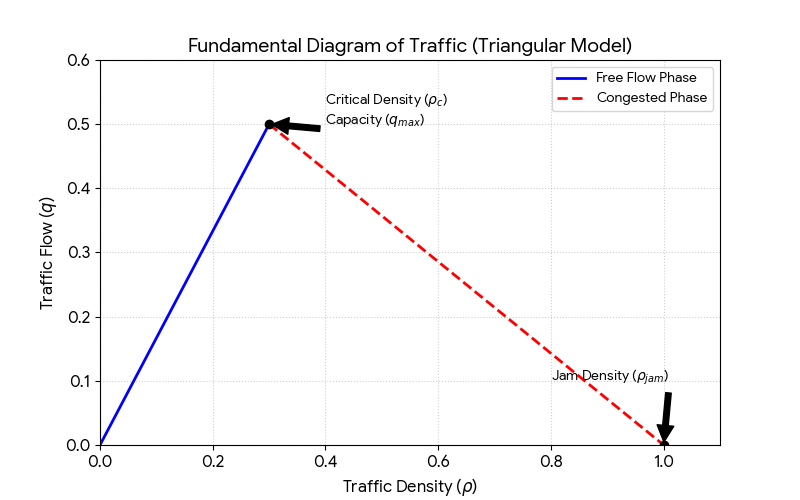}
   %{\color{blue}
   \caption{Schematic representation of the Fundamental Diagram of Traffic (FDT). 
   The triangular shape illustrates the two distinct regimes: 
   the free-flow phase (blue solid line) where flow increases linearly with density,
   and the congested phase (red dashed line) where interactions reduce flow as
   density approaches the jam density ($\rho_{jam}$). The peak represents the road
   capacity ($q_{max}$) at the critical density ($\rho_c$).}
   %}
   \label{fdtgen}
\end{center}
\end{figure}

Vehicular movement in urban networks can be understood like an 
out of equilibrium many-body system with a hierarchy of 
description levels and
relevant scales.
From a microscopic standpoint, traffic is a granular flow
with strong interactions among particles \cite{nobottle}. 
These interactions are
in part
a consequence of decision making of the individual particles themselves, 
which in this sense can be regarded as \emph{agents} which 
can display
non-linear and stochastic behavior. 
At a macroscopic level on the other hand,
traffic can be regarded as a continuous flow in
terms of coarse grained observables like 
densities and average velocities, being perhaps manageable by means of
interventions on road geometry and 
network topology intended to reduce
traffic jams \cite{cassidy, dganzo,loder}. 
Urban traffic capacity is characterized by a 
\emph{Fundamental Diagram of Traffic} (FDT), which indicates the
existence of a free-flow phase and a congested phase with
a transition between them in a critical vehicle density point,
%{\color{blue} 
as further illustrated in the simple triangular 
version of the
flow vs density FDT schematics provided in
the Figure \ref{fdtgen}
%}.
This characteristic FDT has been observed to span several scales
ranging from
the individual street level up to 
aggregate dynamics over the urban network.
At the largest urban scales,
traffic congestion has been shown to display
a remarkable empirical 
correspondence
with epidemics and 
%{ \color{blue} 
contagion propagation
processes, by fitting 
observed congestion to compartmental epidemic models \cite{saberi,duan}.
%}
In the present work
we 
put forward a possible 
unifying framework that explains the scale spanning FDT  
and the epidemic large scale dynamics
by identifying them 
as emergent properties of an out of equilibrium 
system that shares commonalities with 
both granular and continuous
flows. The vicinity of
road intersections are unavoidable sources of 
discontinuity and therefore, the places at which the
granular nature of the flow becomes more apparent.
In other regions far away from the intersections, the continuous density 
description dominates if additional sources
of instability, like bottlenecks, are absent.
Therefore we propose a microscopic contact at the crossroads
mechanism for the transmission of flow in a transport network.
Our resulting model, detailed in the next section, is
conceptually similar to the point queue and traffic flow models 
(see for instance \cite{blubook-vol1-v10} for a discussion on these
approaches)
with the crucial 
difference that the transmission rate parameters are not
defined by 
%{\color{blue}
local characteristics of roads and junctions. 
Moreover, in contrast to previous approaches that
use epidemic models as a metaphor for traffic modeling \cite{sun,saberi,duan}, 
our 
proposed formalism introduced in Section \ref{model}, 
shows
that the mathematical conservation of flux at intersections 
naturally yields the FDT shape. Empirical evidence
for this result is provided in Section \ref{empirical}.
Our first principles derivation of the Fundamental Diagram of Traffic
has theoretical interest on its own and also potential 
applications for traffic control, as discussed in 
Sections \ref{model} and \ref{discussion}. By identifying 
traffic congestion as mathematically equivalent to a transmitted
disease, our model gives possible immunization strategies.
%}

%\subsection{Network flow contact process model}
\section{Mean field network traffic flow model of contagion at the 
crossroads} \label{model}

\begin{figure}[ht]
\begin{center}
   \includegraphics[width=13.0cm]{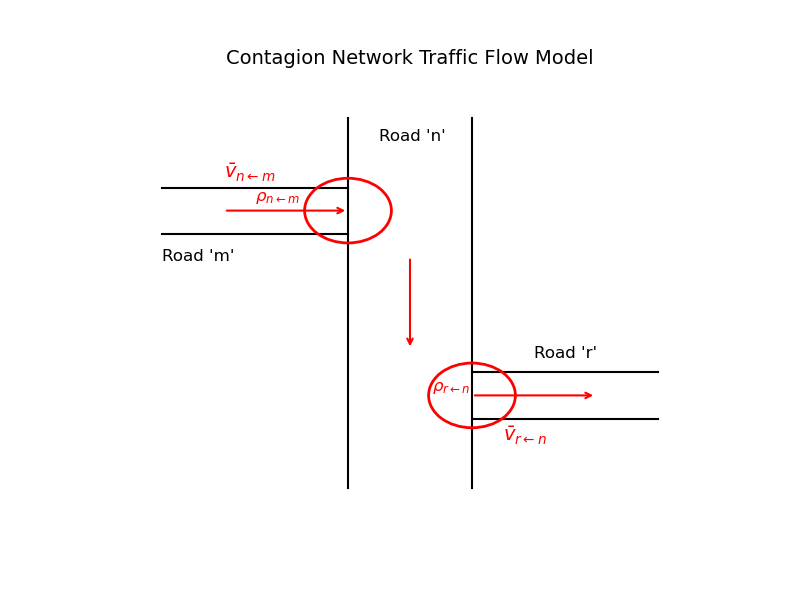}
   %{\color{blue}
   \caption{Junctions level illustration
   of the contagion traffic flow model. 
   In the mean field limit,
   the aggregation of inflow and outflow 
   intersections like those in the 
   Figure, give a coarse grained 
   description equivalent to a 
   Susceptible-Infected-Susceptible (SIS)
   contagion process 
   under stationary flow conditions. If the 
   total transport is constrained to occur 
   in a given time interval, 
   the mean field model is then equivalent to a 
   Susceptible-Infected-Recovered (SIR) process.}
   %}
   \label{pockets}
\end{center}
\end{figure}

Urban traffic has been shown to be
consistent with an out of equilibrium
granular flow process \cite{nobottle}. Consider 
therefore a network
of links (edges), each representing idealized 
roads through which such
a flow occurs. It's additionally assumed 
that  
aggregates of cars 
traversing a link $n$
can be described in terms 
of a continuous car density flow
in any point of the link
except at the
crossroads, which can be regarded as
contact points in which vehicles enter or leave the
$n$-th link.
The master equation, or
instantaneous transition time limit of the Chapman-Kolmogorov equation
\cite{VanKampen}, 
for the car density at an edge $n$ 
is given by,
\begin{eqnarray} \label{ckolmo}
    \frac{d\rho_n(t)}{dt} = \sum_{m\neq n} q_t(n|m) \rho_m(t) -
    \sum_{r\neq n} q_t(r|n) \rho_n(t)
\end{eqnarray}
The $q_t(n|m)$ and $q_t(r|n)$ terms in Eq.(\ref{ckolmo}) represent the 
instantaneous flow rates
of cars from the edges $m$ and $n$ 
to their directly connected  edges $n$ and $r$, respectively.
The precise description of $q_t(n|m)$ and $q_t(r|n)$ 
involve nonlinear interactions of vehicles 
at both sides of a given juncture due to the compromise
between the vehicles in both lanes.
In this sense, traffic signaling
can be understood like a regulator of the inherent instabilities at
street crossroads.
Consider the flows, densities and average speeds in the 
vicinity of the intersections $n \leftarrow m$, where the arrow 
indicates transport from $m$ to $n$, $m\neq n$ 
and at $r \leftarrow n$, $r\neq n$
which vent transport outside the $n$-th road.
In the present work we propose an average description
of the traffic network junctions 
such that pockets of vehicles with 
densities $\rho_{k \leftarrow u}$ are transmitted in an
intersection $k \leftarrow u$ with an average rate of
$\bar{v}_{k \leftarrow u}$, so
\begin{eqnarray} \label{flows}
    q_t(n|m) = \bar{v}_{n \leftarrow m}
    \rho_{n \leftarrow m}(t) \\ \nonumber
    q_t(r|n) = \bar{v}_{r \leftarrow n}\rho_{r \leftarrow n}(t)
\end{eqnarray}
The Eq. (\ref{flows}) can be interpreted like an extension
of the definitional relationship between flow and density
\cite{libre, blubook-vol1-v10}, 
%{\color{blue}
$q = \bar{v}\rho$. Our proposed description broadens the definitional relationship
by considering the $\bar{v}_{k \leftarrow u}$ terms to be the
transport rates of
densely packaged groups of vehicles
at the neighborhood of a given intersection $k \leftarrow u$.
Substituting the Eq. (\ref{flows})
in the master equation Eq. (\ref{ckolmo}),
\begin{eqnarray}\label{mf1}
    \frac{d\langle \rho_n\rangle}{dt} = 
    \sum_{m\neq n} \langle
     \bar{v}_{n \leftarrow m}
    \rho_{n \leftarrow m} \rho_{m}  \rangle
    - 
    \sum_{r\neq n} \langle
    \bar{v}_{r \leftarrow n}\rho_{r \leftarrow n} \rho_n 
    \rangle
\end{eqnarray}
The brackets in Eq. (\ref{mf1}) denote 
spatial averages over the joint density
of vehicles traversing the network. 
In the zeroth order approximation  
of vanishing vehicle density fluctuations,
the mean field dynamics reads,
\begin{eqnarray}\label{mf2}
    \frac{d\langle \rho_n\rangle}{dt} = 
    \langle \rho_{n} \rangle 
    \left [ 
    \sum_{m\neq n} 
    \langle \bar{v}_{n \leftarrow m} \rangle
    \langle \rho_{m} \rangle
    - 
    \sum_{r\neq n} \langle
    \bar{v}_{r \leftarrow n} \rangle
    \langle \rho_{r } \rangle
    \right ]
\end{eqnarray}
The last term inside the square brackets
in Eq. (\ref{mf2}) is simply the venting rate
of vehicles outside the $n$-th link. The first term
inside the brackets
gives the total vehicle transmission 
of incoming vehicles from links other than $n$.
By normalizing spatially over the subgraph
of links directly connected to $n$,
the following coarse grained
dynamics is obtained,
\begin{eqnarray}\label{mfo2}
    \frac{d\rho_n(t)}{dt} = \beta_{n}\rho_{n} (1 - \rho_n) - 
    \gamma_{n}\rho_n
\end{eqnarray}
The mean field expression Eq. (\ref{mfo2}) is
equivalent to a Susceptible-Infected-Susceptible (SIS)
contagion process by identifying 
$\beta_{n}$  and $\gamma_{n}$ like
contagion
and recovery rate parameters, respectively. 
In the present
context, these refer to the effective 
transmission and venting of vehicles to a link $n$ by
its connected network of inwards and outwards 
links.
%}
If a total number of vehicles $C_n$ traverse the $n$-th link 
during a time period $T$, then under
conservative conditions the density $\rho_n(t)$ 
should be normalized in time as,
%{\color{blue}
\begin{eqnarray} \label{constraint}
    \frac{d\rho_n}{dt} = \rho_n  
    \left [(1 - \rho_n)  \beta_n
    - \gamma_n \right ] \quad \text{subject to} 
    \\ \nonumber
    \int_{0}^{T} \rho_n(t) dt = C_n
\end{eqnarray}
The constraints in Eq. (\ref{constraint}) can be satisfied 
by introducing the temporal normalization factors,
\begin{eqnarray} \label{defs}
    I_n \equiv \int_{0}^{T} \rho_{n}(t) dt, \\ \nonumber
    S_n \equiv \int_{0}^{T} (C_n-\rho_{n}) dt \equiv 
    \int_{0}^{T} s_{n}(t)  dt \\ \nonumber
    R_n \equiv C_n - S_n - I_n %\quad C_n = 1 \\ \nonumber
\end{eqnarray}
The Eq. (\ref{defs}) states that
the transport that traverse the $n$-th link during 
a time interval $T$ is conserved.
The introduction of the constraints Eq.(\ref{constraint}) therefore
lead to a 
Susceptible-Infected-Recovered (SIR)
type model for the
traffic flow over the $n$-th node, in which
a time span $T$ with total transport $C_n$
can be interpreted as an individual
contagion process.
By combining with the vanishing 
spatial density fluctuations
limit of Eq. (\ref{mfo2}), the following 
coarse grained SIR description of an individual link
in a mean field network is obtained,
\begin{eqnarray} \label{sisf}
   s(t) &=&  
    s(0) e^{- \beta \rho(t)} \\ \nonumber
   \frac{d\rho(t)}{dt} &=& \beta
   \rho(t) s(t) - \gamma \rho(t)
\end{eqnarray}
%}

\subsection{Comparison with related approaches and the Fundamental 
Diagram of Traffic} \label{contrib}

Descriptions of traffic based on contagion dynamics models have been previously 
proposed, but these depend on ad-hoc definitions of congested states. For instance,
in the seminal work \cite{sun}, congestion is defined by mapping to bond percolation
on a small world network, where the congested states 
can be transmitted according to a contagion rate parameter and a percolation 
probability parameter. More recently in \cite{saberi}, congested states are
formalized in terms of a threshold for the instantaneous average speed
on a link relative to its maximum speed. Given this definition
for a congested state road, 
is then proposed in \cite{saberi} that congestion 
spreads through the network by a mechanism that at the aggregated level
is described by a simple compartmental SIR dynamics. 
Ample empirical evidence is given in \cite{saberi}, that indicates that the
compartmental SIR 
behavior closely match the temporal evolution of congestion in 
a given city.

In our model, congestion
is not treated like a state of a link or group of links, but rather
emerges in the form of a shape for flow vs density 
given a total transport in a given time. From 
Eq. (\ref{sisf}) it follows that  
(with $s_0 \equiv s(0)$),
$\dot{\rho}(\rho)$ is expressed by,
\begin{eqnarray} \label{fdt}
   \dot{\rho}(\rho) \approx
   \beta
   \rho s_0e^{- \beta \rho} - \gamma \rho
\end{eqnarray}
The resulting shape is consistent with the 
Fundamental Diagram of Traffic (FDT) \cite{blubook-vol1-v10}, 
in the sense that it permits to distinguish between free flow and congested 
phases by the sign of the derivative of the flow with respect to the density.

From Eq. (\ref{fdt}), it follows that the macroscopic
critical density point is approximately given by the
solution of the transcendental equation,
\begin{eqnarray} \label{critical}
   \rho_c = 1 - 
   \frac{\gamma}{\beta s_0}e^{\beta \rho_c}
\end{eqnarray}

%{\color{blue}

The result Eq. (\ref{critical}) clearly expresses 
the value of our mean field approach for the
understanding of traffic in terms of contagion
dynamics.
Previous epidemic models of road networks, 
like the one presented in \cite{saberi},
are largely reactive: they track the spread 
of a jam after certain speed threshold is breached. 
Our result Eq. (\ref{critical}) in contrast, 
offers a predictive capability. It implies that the ``tipping point''
into congestion is not a universal constant but is a function of
the inflow or infection rate ($\beta$), the 
outflow or recovery rate ($\gamma$) and 
$s_0$ (the network load).
This suggests that a traffic controller could theoretically 
increase the critical density 
(making the road more resistant to jams) by altering $\gamma$
(e.g., adjusting traffic light cycles to increase venting) relative to $\beta$. 
This provides a theoretical basis 
for ``immunizing'' intersections against jams 
by tuning their topological parameters, 
rather than just managing queues once they form.
%}

\section{Comparisons with empirical data} \label{empirical}

\subsection{Direct experimental evidence} \label{saltillo}

To have data for the direct empirical evidence 
of the proposed mean field model 
at the single link level requires precise flow measurements 
from all the relevant intersections
in an entire subgraph. This can be difficult due to the 
usually incomplete coverage provided by 
traffic sensors. Here however 
we present an empirical study that intends to 
overcome the stated obstacle by analyzing vehicle flow
data in a lane segment close 
to an intersection where a group of important 
traffic links meet. The case study 
is situated in the city of Saltillo, capital of the state of
Coahuila in Mexico. The data has been
gently provided by Autonomous University of
Coahuila professor
Dr. Jaime Burgos García together with the 
Municipal Institute of Sustainable Urban Mobility of Saltillo. 
The experimental setup has been implemented as follows.
A vehicle count station 
for the road segment of interest, which is part of the
Venustiano Carranza Boulevard in Saltillo, 
has been situated 
at the approximate
latitude and longitude coordinates 
$(25.459628, -100.983199)$. 
Another vehicle count station 
has been placed nearly $1.0$ Km in direction
north, at the approximate coordinates 
$(25.466904, -100.979632)$. The second station
monitors the already mentioned 
hub of lanes, one of those corresponding to
the chosen road section of the Venustiano Carranza Boulevard. 
By the combined vehicle count measurements of both stations, 
which haven taken in $15$ minutes time intervals, the
number of cars
entering and leaving the road segment through the hub and
the total number of vehicles in the segment have been registered.
The aggregated data is
presented in the Table \ref{tab:1}, in which the 
number of vehicles traversing the lane segment at the hub intersection
is calculated as the difference between
the total input plus output vehicles and the number of vehicles
incoming from the 
Venustiano Carranza Boulevard road segment itself and continuing
their transit along it.
A total of $20$ measurements have been
made the $25$th of September, $2024$. The first ten measurements have been 
taken at the morning traffic peak, 
which goes from $7:00$ AM to $9:30$ AM, while the 
second half of the dataset instances correspond to the afternoon peak
hour, from $17:00$ PM to $19:30$ PM, local times. 

\begin{figure}[ht]
\begin{center}
   \includegraphics[width=13.0cm]{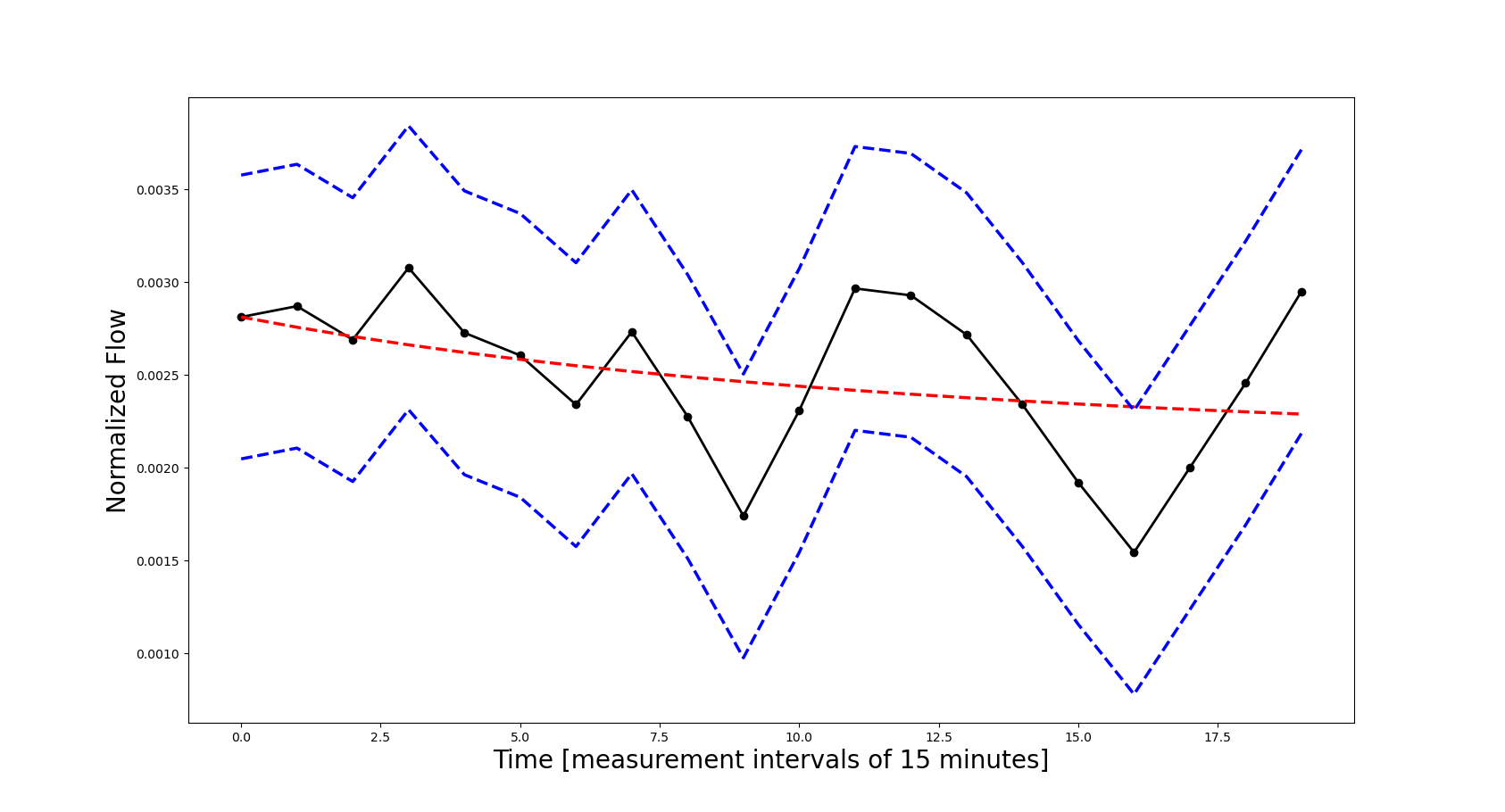}
   \caption{Fitting of the SIR mean field model to
   experimental data in the city of Saltillo, Mexico
   (see Subsection \ref{saltillo} for a thorough explanation).}
   \label{Saltillo1}
\end{center}
\end{figure}

\begin{figure}[ht]
\begin{center}
   \includegraphics[width=13.0cm]{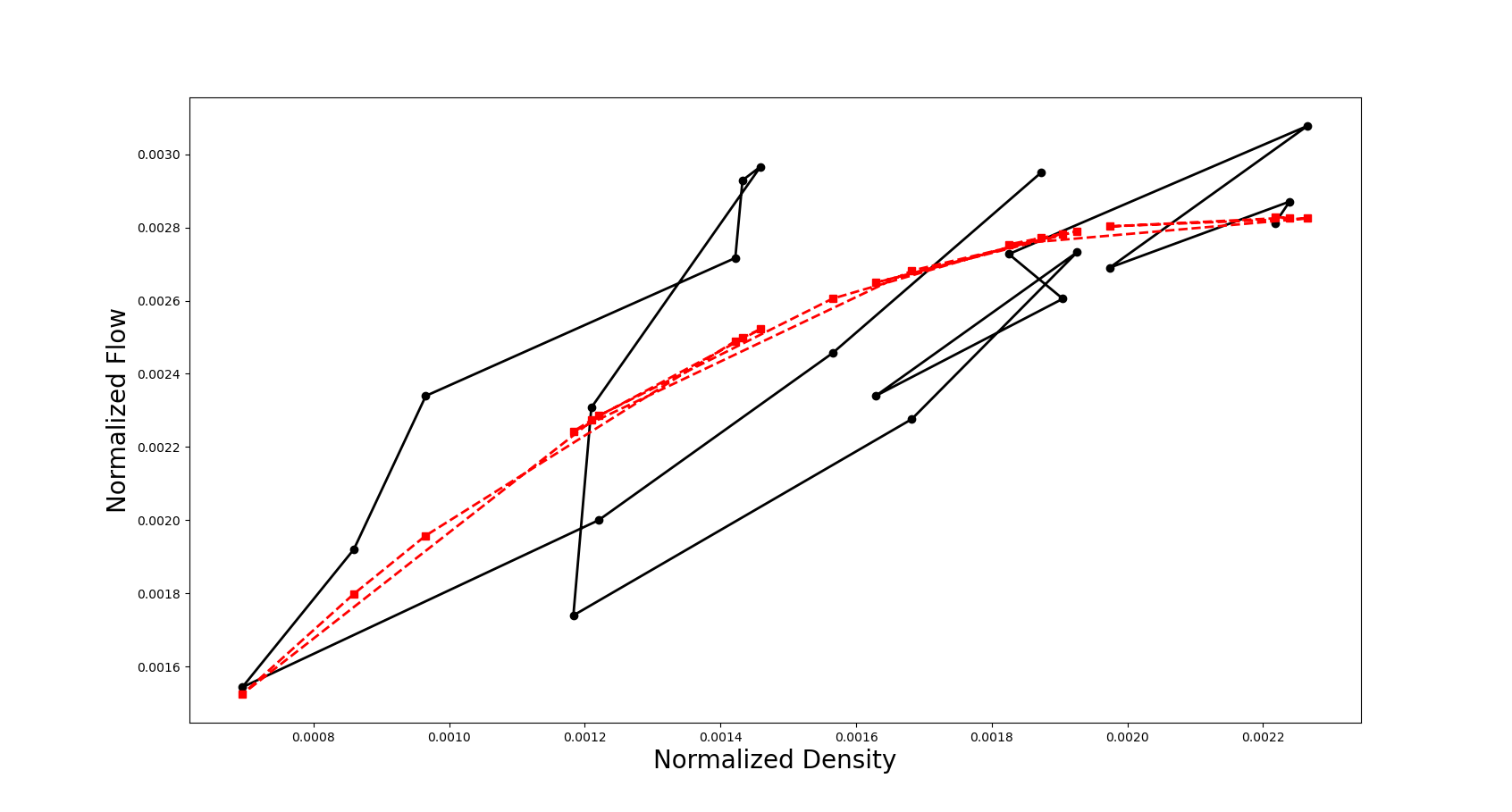}
   \caption{Experimental vehicle flow vs density data from
   the city of Saltillo, Mexico (black dotted lines) and the
   FDT function that results from the estimated SIR model
   (red dotted dashed lines).}
   \label{Saltillo2}
\end{center}
\end{figure}

We have used the provided data to test our mean field contagion model
by normalizing the vehicles counts per measurement considering the
total transport through the intersection during the total observation 
time span. Therefore the car density flux at the intersection is normalized 
as,
\begin{eqnarray}
   \Delta \rho/\Delta t = 
   \frac{(\text{number of vehicles traversing the intersection})}
   {(20*\text{total transport}) }
\end{eqnarray}
In this way, the data in the last column 
of the Table \ref{tab:1} 
gives a vehicle flux time series that
we have used to fit the discrete Euler
approximation of the SIR model, 
Eq.(\ref{sisf}), neglecting fluctuations.
The results are shown in the Figure \ref{Saltillo1},
where the dotted lines represent the flow data, the
dashed blue lines its $90$\% confidence interval 
and the middle dashed red line gives the fitted 
SIR model. The predicted average rate parameters
turn out to be 
$\beta = 24.435 \pm 4.0$, 
$\gamma = 23.205 \pm 4.0$, which are consistent
with the actual observed rates of $25$ and $27$ 
vehicles per minute, respectively.
Figure \ref{Saltillo2}
on the other hand, uses the predicted
mean
$\beta$ and $\gamma$ rates to evaluate
the function $\dot{\rho}(\rho)$, which 
gives the Fundamental Diagram of Traffic (FDT)
for the considered road segment at the 
considered time spans. 
%{\color{blue} 
The large confidence 
intervals shown in Figure \ref{Saltillo1}
confirm that significant fluctuations are present, 
which naturally leads to scatter in the FDT representation of 
Figure \ref{Saltillo2}.
Besides the statistical noise resulting from the 
limited sample size,
the empirical data also displays the phenomenon of
hysteresis loop in the Fundamental Diagram, which is
considered to be a consequence 
of drivers behavior during acceleration and deceleration
\cite{traffhys}. It's interesting to notice that our model
successfully identifies 
the attractor (the red line) 
around which the real, noisy, 
hysteretic data (the black line) oscillate. 
That is, our mean field model  
successfully captures the central tendency 
(the ``backbone'') of the Fundamental Diagram, 
effectively averaging through 
microscopic interactions like those resulting in
the hysteresis loop of the experimental data.
%}

%Please add the following packages if necessary:
%\usepackage{booktabs, multirow} % for borders and merged ranges
%\usepackage{soul}% for underlines
%\usepackage{xcolor,colortbl} % for cell colors
%\usepackage{changepage,threeparttable} % for wide tables
%If the table is too wide, replace \begin{table}[!htp]...\end{table} with
%\begin{adjustwidth}{-2.5 cm}{-2.5 cm}\centering\begin{threeparttable}[!htb]...\end{threeparttable}\end{adjustwidth}
\begin{table}[!htp]\centering
\caption{Vehicles count data 
at the peak hours in the vicinity of a
highways intersection in Saltillo, Mexico}\label{tab:1}
%\resizebox{ extwidth}{!}{ % use this if the table is too large
\begin{tabular}{lrrrrr}\toprule
15 minutes intervals &Input vehicles 
&Output vehicles &Vehicles in the &Vehicles traversing\\\cmidrule{1-5}
& & &segment of link &the intersection \\\cmidrule{4-5}
1 &466 &482 &418 &530 \\\cmidrule{1-5}
2 &502 &461 &422 &541 \\\cmidrule{1-5}
3 &444 &435 &372 &507 \\\cmidrule{1-5}
4 &521 &486 &427 &580 \\\cmidrule{1-5}
5 &445 &413 &344 &514 \\\cmidrule{1-5}
6 &428 &422 &359 &491 \\\cmidrule{1-5}
7 &364 &384 &307 &441 \\\cmidrule{1-5}
8 &454 &424 &363 &515 \\\cmidrule{1-5}
9 &373 &373 &317 &429 \\\cmidrule{1-5}
10 &256 &295 &223 &328 \\\cmidrule{1-5}
11 &294 &369 &228 &435 \\\cmidrule{1-5}
12 &373 &461 &275 &559 \\\cmidrule{1-5}
13 &356 &466 &270 &552 \\\cmidrule{1-5}
14 &334 &446 &268 &512 \\\cmidrule{1-5}
15 &253 &370 &182 &441 \\\cmidrule{1-5}
16 &221 &303 &162 &362 \\\cmidrule{1-5}
17 &204 &218 &131 &291 \\\cmidrule{1-5}
18 &284 &323 &230 &377 \\\cmidrule{1-5}
19 &343 &415 &295 &463 \\\cmidrule{1-5}
20 &443 &466 &353 &556 \\\cmidrule{1-5}
& & & & \\
Average rates per minute &{\bf 25}
&{\bf 27}  & & \\\cmidrule{1-3}
Total transport & & & &{\bf 9424}\\\midrule
\bottomrule
\end{tabular}
\end{table}

\subsection{Comparisons with traffic sensors data} \label{sensors}

Our contagion network traffic flow 
model has also been tested on
representative instances of the publicly available 
UTD19 database \cite{loder}. This database provides 
fundamental traffic variables including
flow and density, from 41 cities worldwide.
The data is collected from stationary traffic sensors,
most of them inductive loop detectors,
which report traffic flow 
and the fraction of time during an observation period in
which a detector is occupied as a proxy
for the vehicle density. 

\begin{figure}[ht]
\begin{center}
   \includegraphics[width=13.0cm]{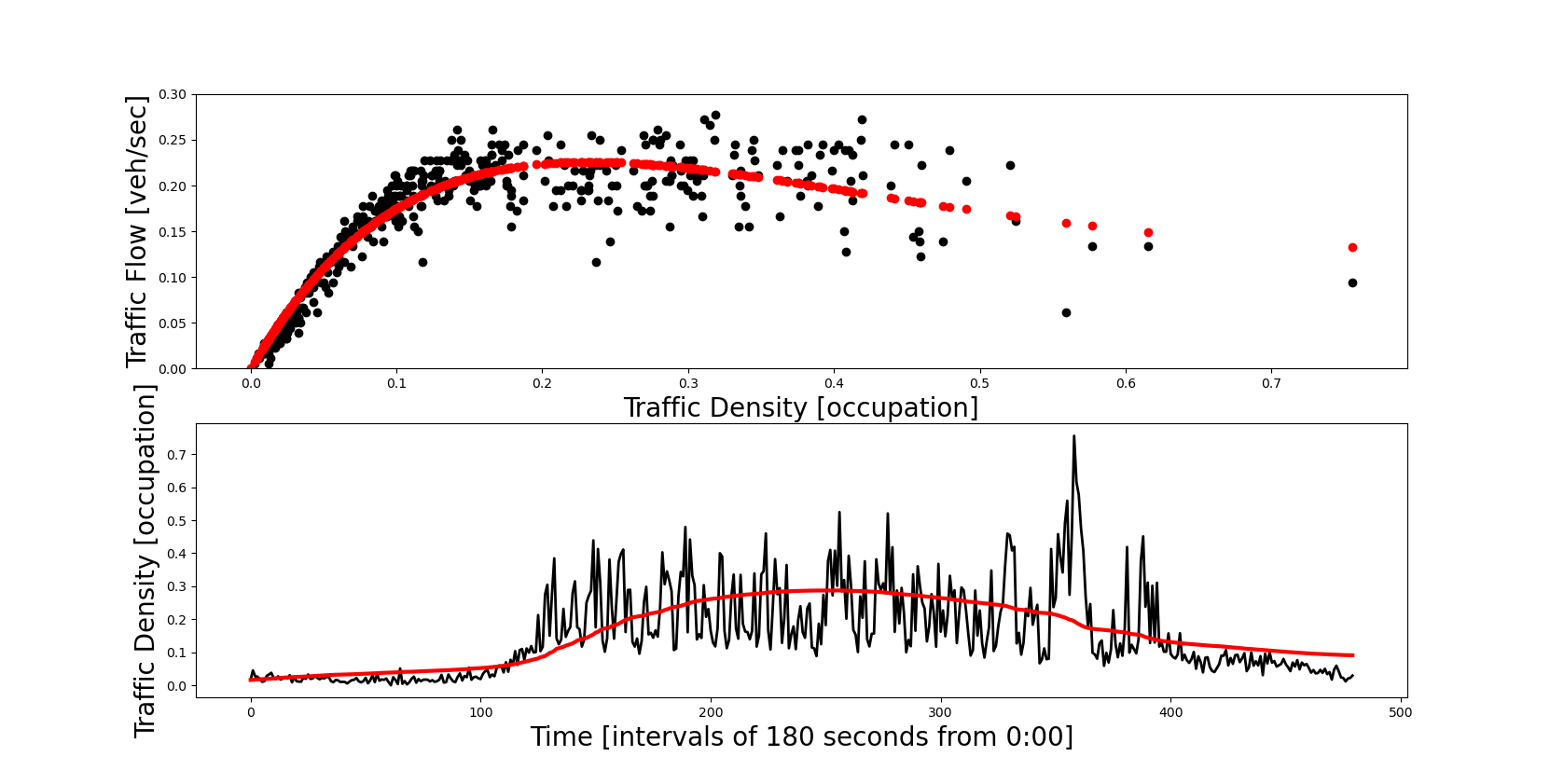}
   \caption{Traffic dynamics from a single detector of
   the Schimmelstrasse street in the
   Aussersihl district of Zurich, Switzerland, at 
   Wednesday 28th October, 2015. 
   %In this and 
   %in all of the following graphs, 
   Empirical data is shown in black while
   the fit
   of the local mean field
   contagion model to the data is shown in red.
   }
   \label{Aussersihl0}
\end{center}
\end{figure}

The upper and
lower panels of the Figure \ref{Aussersihl0},
show
the fitting of the SIR density dynamics from a 
single detector over a particular
street, in the Aussersihl district of 
Zurich, Switzerland, over a $24$ hour period. 
This instance 
of the UTD19 database 
has been showcased in a number of works 
and has good quality sensor data \cite{loder}.  
Fitting of the SIR rate 
parameters is done by minimizing
the total squared error between the observed densities
and those predicted by the SIR 
dynamics integrated by first 
order discretization through the Euler method.
Because the street density data is a proxy,
additional parameters for a linear transformation   
that resolves the particular sensor calibration are introduced.
The resulting optimization
problem has two levels. Firstly the rate parameters are
estimated by solving
\begin{eqnarray} \label{op1}
    \textit{min} _{\beta_n, \gamma_n} \quad \sum_{t=1}^{T} 
    [\hat{\rho}_n(\beta_n, \gamma_n, t) - \rho_n(t) ]^ 2 
\end{eqnarray}

In Eq.(\ref{op1}), $\hat{\rho}_n$ is the 
vehicle density in the $n$-th link given by
the one step ahead prediction
of the SIR model at 
the observed time $t$.
The sum 
of squared errors is calculated from the total number
of observations $T$, over the $24$ hour period.
The local fitting of the SIR dynamics has been done
using the Broyden–Fletcher–Goldfarb–Shanno algorithm,
assuming time independent rate parameters, 
$\beta_n(t) = \beta_n$, $\gamma_n(t) = \gamma_n$
and also assuming that
$s(0) = 1$.

In a second stage, optimal calibration parameters $s_0, c_1, c_2$ are
calculated by,
\begin{eqnarray} \label{op2}
    \textit{min} _{s_0,c_1,c_2} \quad \sum_{t=1}^{T} 
    \left [\hat{y}_n(s_0,c_1,c_2,t) - \frac{d\rho_n(t)}{dt} \right ]^ 2 
    \\ \nonumber
    \hat{y}_n(s_0,c_1,c_2,t) \equiv s_0 \beta_n \rho_n(t) e^{-\beta_n \rho_n(t)}
    - \gamma_n \rho_n(t)
\end{eqnarray}

The calibration parameter $0<s_0<1$ estimates the fraction of the overall
transport covered by the $n$-th link in the mean field limit. The 
parameters $c_1$ and $c_2$ are mere calibration constants for the rates,
$\beta_n \to c_1 \beta_n$, $\gamma_n \to c_2 \gamma_n$.

\begin{figure}[ht]
\begin{center}
   \includegraphics[width=13.0cm]{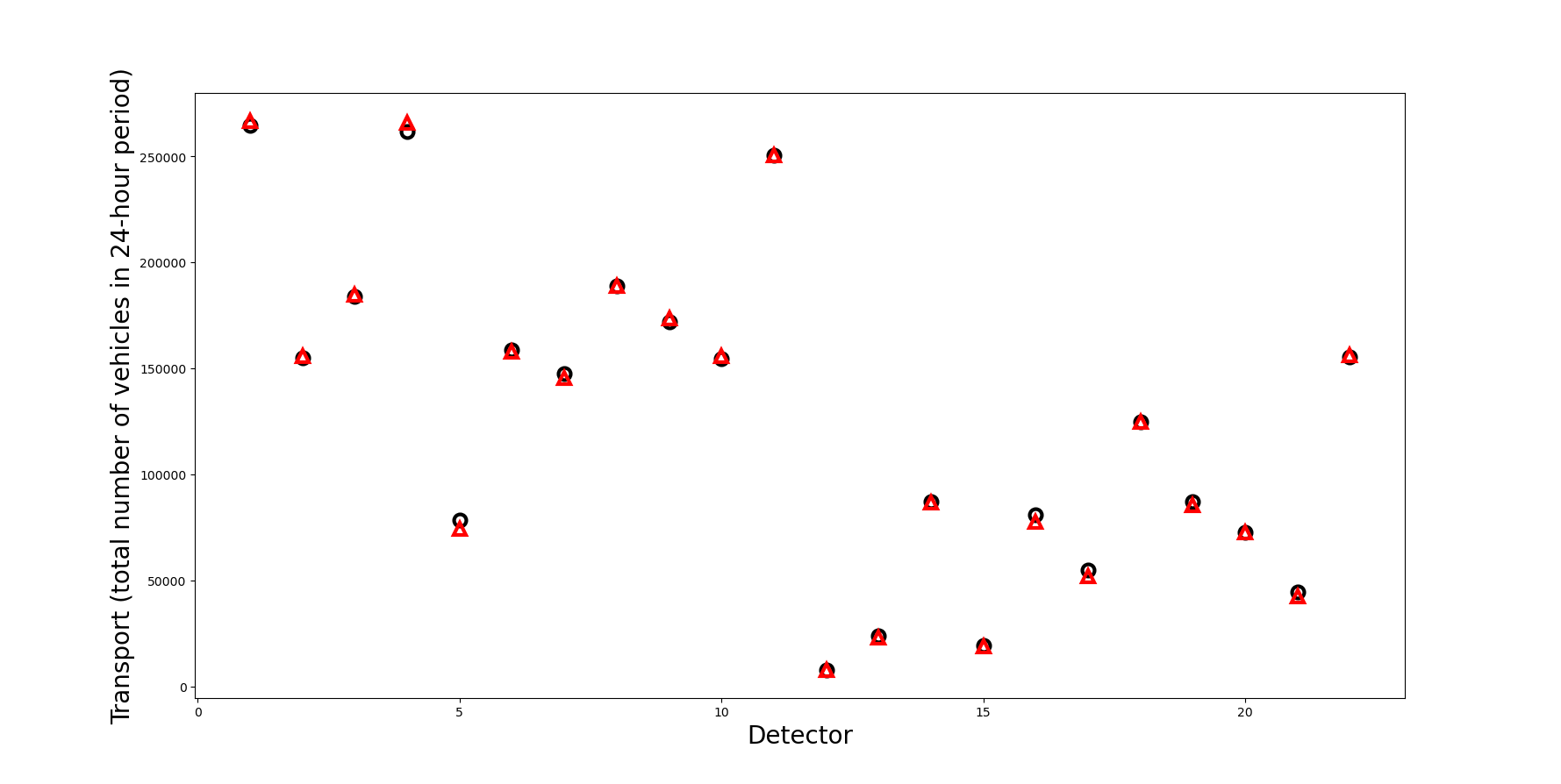}
   \caption{Predicted (red triangles) 
   and observed (black circles) transport in 22 detectors
   from the Aussersihl district in Zurich, Switzerland, at
   Wednesday 28th October of 2015.
   From the estimated transmission rate parameters, the mean field
   crossroad contagion network traffic flow model predicts 
   an average 
   %{\color{blue} 
   inflow over
   the covered region of $30.92$ veh/hr and an
   average venting rate of $6.54$ veh/hr.
   %}
   The arbitraily chosen detectors from the UTD19 database are,
   K14D15, K14D13, K14D14, K17D13, K17D14, K17D15, K17D12, K17D11,
   K13D14, K13D13, K13D12, K14D12, K11D19, K11D11, K12D16, K133D11,
   K133D12, K133D13, K11D12, K11D14, K12D14, K12D17.}
   \label{Aussersihl1}
\end{center}
\end{figure}

The upper panel of the Figure \ref{Aussersihl0} displays the estimated
and observed
FDT, $\hat{y}_n (\rho_n)$ and $\dot{\rho}_n (\rho_n)$ respectively. 
The lower panel on the other hand, shows the estimated and observed 
SIR temporal dynamics of the vehicle density. 
Figure \ref{Aussersihl1} reports the comparison between
the predicted and observed total transport in the $24$ hour period
for an arbitrary set of detectors from links in the vicinity and 
same district of the detector reported in the Figure \ref{Aussersihl1}.
The links rate parameters are estimated 
by the previously discussed procedure individually. 
The estimated rate parameters averaged over the group of links
are consistent with what expected 
from the equation (\ref{sisf}).

\section{Discussion} \label{discussion}

Despite being an idealization, the contagion framework to 
traffic appears to capture features 
that are
essential to characterize urban 
transport at different spatial and temporal
scales.
The formalism integrates granular with continuous aspects of 
vehicle flow in conjunction with basic temporal 
constraints, leading to a local mean field contagion type
model. Our empirical comparisons indicate
that single detector data aggregates information from all of 
the links directly connected to the detector's road.
In fact, the experiment presented in Subsection \ref{saltillo}
indicates that it's possible to estimate macroscopic traffic network
parameters, like total transport or optimal vehicle transit
rates over a region, from purely local flow measurements at
an intersection.

The macroscopic mean field equation (\ref{sisf}), 
predicts that
aggregating links in a given 
macroscopic area of the roads network should converge to
macroscopic rate parameters that reflect the 
vehicle flow in the corresponding network's section.
This appears to be consistent with the empirical
study reported in the Figure \ref{Aussersihl1}. 

Clearly a next important step in the development of 
the proposed mean field theory,
is to apply our approach 
to large macroscopic traffic networks datasets, by carefully
taking into account the network topology. 
Our equation (\ref{critical}) indicates that 
the free flow
optimal state of a macroscopic 
traffic network occurs in
a critical density given in terms of
aggregated rate parameters $\beta$, $\gamma$ and a 
normalization parameter associated with the total
transport in the given time period, $s_0$.
%{\color{blue}
This may lead to ``immunization'' traffic control strategies, as
explained in Subsection \ref{contrib}.
%}

Generalizations to the mean field setup that consider network's 
spatiotemporal structures in greater detail, can
in principle be explored 
by taking advantage of recent 
developments in the study of contagion models on networks 
with complex topologies, like for 
instance those presented in \cite{palafox}.

\section{Acknowledgments}

The authors are grateful with Dr. Jaime Burgos García 
and Dr. Simón Rodríguez Rodríguez
for the very useful discussions with them
and particularly to Dr. Burgos García for 
having kindly shared with us the 
experimental data from the 
Municipal Institute of Sustainable Urban Mobility of Saltillo
analyzed by us in Subsection \ref{saltillo}.

The authors also acknowledge to the creators and maintainers of the 
UTD19 database, from which the traffic sensors data used in
Subsection \ref{sensors} has been taken. 
The UTD19 database is hosted in the site utd19.ethz.ch

\newpage

\bibliographystyle{unsrtnat}
\bibliography{ref}

\end{document}